# Ethical considerations for data involving human gender and sex variables

Suzanne Thornton [*]     Rochelle E. Tractenberg [†]

Abstract

The inclusion of human sex and gender data in statistical analysis creates a nuanced problem for the modern statistics practitioner. There are multiple, overlapping, considerations for data collection, combination, analysis, and interpretation that involve variables intended to represent sex and/or gender. A few of these considerations include the privacy, protection and ethical treatment of human subjects, avoiding non-response and missingness, and mitigating the potential for classification bias. These considerations are not unique to variables representing sex and gender, nor to decisions that are informed by analyses involving these variables. However, consideration of the relevance of aspects of the ethical practice standards for statistics and data science to sex and gender variables is timely, with results that can be applied to other socio-cultural variable definitions and selection. Historically, human gender and sex have both been categorized according to a binary system. The continuation of this tradition in modern research persists mainly because it is easy, and not because it produces the most valuable scientific information. Utilizing the traditional binary classification of gender enables older and newer data sets to be wrangled, munged, and combined easily. However, this classification system eliminates the individual's ability to articulate and assert their gender identity and conflates gender and sex. Moreover, defaulting to a binary system for sex obscures potentially important differences by collapsing across many valid and authentic categories (e.g. involving primary and/or secondary sex characteristics). This approach perpetuates historical, inaccurate, simplicity - and bias - while also limiting the information that emerges from analyses of human data. These limitations also violate multiple elements in the American Statistical Association's (ASA) Ethical Guidelines for Statistical Practice. Information that would be captured with a non-binary classification (including the prevalence of differences in sexual development) could be relevant to decisions about analysis methods as well as to decisions based on otherwise expert statistical work. Modern statistical practitioners are increasingly concerned with inconsistent, uninformative, and even unethical data collection and analysis practices. This paper presents a historical introduction to the collection and analysis of human gender and sex data, offers a critique of a few common survey questioning methods based on alignment with the ASA Ethical Guidelines, and considers the scope of ethical considerations for human gender and sex data from the early stages of design through the later stages of analysis.

## 1. History of gender and sex data

Social and biomedical sciences have recently begun to address the fact that traditional binary, and even categorical, treatment of sex and gender do not sufficiently capture either the human experience or the true effect of variation in social, psycho-social, or biomedical studies.(Beery and Zucker, 2011; Holdcroft, 2007; McCarthy et al., 2012; McGregor et al., 2016; Miller et al., 2011) The use of categorical variables in statistics and data science -in any analysis where probabilities are assigned- assumes that the categories are mutually exclusive and exhaustive (Larsen and Marx, 1986). While it can be a simple matter to assign an individual to either one or another of traditional sex or gender categories and follow research guidelines, say, those set forth by the Sex and Gender Equity in Research group (Heidari et al., 2016), such assignments may not be accurate. Traditional gender and sex

[*]Department of Mathematics and Statistics, Swarthmore College, 500 College Ave, Swarthmore, PA 19081

[†]Collaborative for Research on Outcomes and -Metrics, and Georgetown University, Building D, Suite 207, 4000 Reservoir Road NW, Washington, DC 20057

assignments can be inaccurate in terms of which category is chosen, say a person belongs to more than one category (i.e., the categories are erroneously characterized as "mutually exclusive"), or because a person belongs to another category that is not considered e.g., Harrison et al. (2012). Any of these inaccuracies can create bias, imprecision, and errors in judgments or decisions based on the analyses of such data which calls for a critical statistical assessment.

Sex is typically defined based on phenotype (anatomical or chromosomal patterns see, e.g., Wackwitz (2003) and Ritchie (2003)), while gender is typically defined based on personal representation of socially constructed norms relating to sex, e.g., in terms of dress and behaviors. Historically, in western cultures, norms for anatomic and social categorization have been "male"/"masculine" and "female"/"feminine". Unfortunately, the vast majority of research produced by the global west has ignored human variation in development and expression by conflating these two variables, treating sex and gender as synonymous - which reflects the alignment of biological and social statuses (i.e., male with masculine; female with feminine). While this alignment might yield the correct use of a categorical variable for the majority of (western) individuals, it is clearly incorrect for anyone whose sex and/or identity are either not fixed (e.g. gender-fluid) or otherwise, not compatible with definitions of "categorical" that are crucial for mathematical and probabilistic properties to pertain. Collapsing across the many differences in human variation might be convenient, but it is incorrect and often inappropriate. The cautious practitioner (data scientist, data analyst working with scientists, or scientist doing their own data science/statistical analysis) must recognize, and seek to move beyond, the assumptions of traditional practice. A careful consideration of the accuracy and true utility of variables that purport to capture or reflect gender and sex characteristics exposes the powerful role of measurement assumptions, variable selection, and model building on decisions or conclusions based on statistical analyses and data science. As such, these considerations are also highly relevant for variables like "race" or "ethnicity", and even "socioeconomic status".

Beyond the potential implications for probability estimation from incorrect classification of categories as "mutually exclusive and exhaustive", the most obvious problem with utilizing an inappropriate binary classification scheme is the propagation of estimation errors that arise from assuming a binomial probability distribution instead of a multinomial distribution. A multinomial model specifies the probability ($\pi_j$) for each of the $j$ possible outcomes and counts the number of individuals in each category. Incorrect model choice results in an incorrect expected value (binomial: mean ($\pi$), multinomial: mean ($n\pi_j$)) and an incorrect variance that is incorrect by an unpredictable amount, depending on each category's mean, whenever the the number of individuals is unknown (binomial variance: ($\pi(1 - \pi)$); multinomial variance: $n(\pi_j(1 - \pi_j))$) (Agresti, 2003). Knowingly utilizing incorrect distributional assumptions violates multiple ASA Ethical Guidelines (presented in the Appendix), e.g., A2, "(the ethical statistical practitioner) Uses methodology and data that are valid, relevant, and appropriate, without favoritism or prejudice, and in a manner intended to produce valid, interpretable, and reproducible results.", and Principle B: "The ethical statistical practitioner seeks to understand and mitigate known or suspected limitations, defects, or biases in the data or methods and communicates potential impacts on the interpretation, conclusions, recommendations, decisions, or other results of statistical practices."

Statisticians are not typically expected to tell the scientists or other collaborators with whom they work what needs to be measured. However, the statistician is ideally positioned to advise researchers in a way that encourages better measurement practices - practices that assure whatever is being assessed both captures the characteristics of interest to the scientist and aligns with the types of analyses that are proposed or envisioned (e.g., ensuring methods for interval, nominal, or ordinal data are utilized and that categories are not ignored when they could affect results and their interpretability). This is an important role statisticians play in the advancement of science. Moreover, the ASA Ethical Guidelines articulate the ethical practitioner's obligations to: "Communicate data sources and fitness for use, including data generation and collection processes and known biases" (B1); to be "transparent about assumptions made in the execution and interpretation of statistical practices including methods used, limitations, possible sources of error, and algorithmic biases" (B2); and to "Consider the impact of statistical practice on society, groups, and individuals. Recognize that statistical practice could adversely affect groups or the public perception of groups, including marginalized groups. Consider approaches to minimize negative impacts in applications or in framing results in reporting" (D6). In fact, among the responsibilities that practitioners have to research subjects, data subjects, or those directly affected by statistical practices (Principle D), the ethical statistical practitioner "understands the provenance of the data, including origins, revisions, and any restrictions on usage, and fitness for use prior to conducting statistical practices" (D10). Moreover, the ethical practitioner "avoids compromising validity for expediency. Regardless of pressure on or within the team, does not use inappropriate statistical practices" (E4) and "serves as an ambassador for statistical practice by promoting thoughtful choices about data acquisition, analytic procedures, and data structures among non-practitioners and students." (F5). These points emphasize the diversity of obligations accruing to the statistical practitioner that support specific attention to variable definition and selection. As mentioned previously, this is not limited to variables representing sex and gender only, but also reflects the importance of effective interdisciplinary communication about measurement and meaning in variable selection for statistical analysis.

## 2. Gender and sex minorities in (modern) ethical research design

A 2022 article in Significance addresses the statistical practitioner audience in calling for a more inclusive – and informative – approach to collecting data on human gender and sex.(Thornton et al., 2022) The authors address three recommendations:

1. Ensure measurement accuracy and specificity.

2. Plan (and act) to protect study participants' privacy.

3. Center the inclusivity and respect of all research participants.

The first recommendation is to ensure measurement accuracy and specificity through identifying what information is relevant for the question(s) to be addressed with the data collection and/or analysis. This recommendation is relevant for both the data curator/collector and the data analyst: both must carefully define and select relevant variables for a particular statistical model to assure that it will generate actionable results relevant to the question. Some social scientists have opted to use scales for masculinity and femininity (e.g., Kachel et al. (2016)) as an attempt to emphasize when gender (rather than sex) is the variable of interest. There are, in fact, a multitude of such scales, so this approach still requires careful consideration of what is meant by masculinity and femininity, what their roles are in the target research question, and how to communicate these concepts to data contributors with labels that are independent of cultural, age, and personal differences in perception.

There is no easy solution, even in a biomedical context, to identify which sex features are relevant characteristics for a study. A 2018 article in Endocrine Reviews summarizes the underlying concern with conflating terms associated with gender and sex in noting that "the rigor of research depends on researchers' understanding of the ways in which sex and gender influence the biologic systems they study... Categories of sex include males, females, intersex individuals born with male and female characteristics, and people who undergo interventions to reassign their sex. In some instances, syndromes resulting from atypical sexual development can complicate categorization of sex."(Rich-Edwards et al., 2018, pp.431–432) Fortunately, the modern research landscape seems to be responding to the challenge, for example, many Nature Potforlio journals now prompt researchers submitting articles to "state whether and how sex and gender were considered in their study design, or to indicate that no sex and gender analyses were carried out, and clarify why."(NAT, 2022, pp.1)

In addition to serving the purposes and objectives of the data curation/collection and analysis, ensuring measurement accuracy and specificity also brings the statistical practitioner in line with many ASA Guideline elements across multiple Guideline Principles (e.g., see principles A2, A3, A4; B, B1, B2, B6, B7; C1, C4, C8; D, D2, D6, D10; E4; and F5 in the Appendix). Although it might be considered to be most relevant to data collection/curation or the planning/design phase of a project, the recommendation to ensure measurement accuracy and specificity has implications for selection of methods, interpretability of results, and the reproducibility of those results.

The second recommendation from the Significance article is to plan to protect study participants' privacy at every stage possible from data collection, to analysis, to data storage for future use. This recommendation is consistent with federal regulations as well as core bioethics principles respecting every individual's basic human right to autonomy with respect to their personal health information. In the United States, the Privacy Rule protects all "individually identifiable health information" held or transmitted by a covered entity or its business associate, "in any form or media, whether electronic, paper, or oral." The Privacy Rule calls this information "protected health information (PHI)" (DHHS, 2003) and "[h]ealth information such as diagnoses, treatment information, medical test results, and prescription information are considered protected health information under HIPAA, as are national identification numbers and demographic information such as birth dates, gender, ethnicity, and contact and emergency contact information."(HIPPA, 2022) Note that gender is considered PHI, but sex is not in the HHS documentation. However, sex and gender are both considered "personally identifiable information" (PII), particularly when used with other information. Just as it is complicated to categorize sex (Rich-Edwards et al., 2018) it is also challenging to identify when an individual's sex and gender data are to be considered PHI, PII, or both. The ASA Ethical Guidelines state that the ethical statistical practitioner "keeps informed about and adheres to applicable rules, approvals, and guidelines for the protection and welfare of human and animal subjects. Knows when work requires ethical review and oversight."(D1) Because of the complexities of correctly determining whether sex and gender variables are protected as PHI, PII, or both, this Guideline element is especially relevant to both this recommendation and to the previous one. Effective interdisciplinary communication is critical to assure that data to be curated or collected will be protected correctly. Whether collecting/curating or analyzing data, the ethical practitioner "Knows the legal limitations on privacy and confidentiality assurances and does not over-promise or assume legal privacy and confidentiality protections where they may not apply." (D9); "Understands the provenance of the data, including origins, revisions, and any restrictions on usage, and fitness for use prior to conducting statistical practices." (D10); and "Does not conduct statistical practice that could reasonably be interpreted by subjects as sanctioning a violation of their rights. Seeks to use statistical practices to promote the just and impartial treatment of all individuals." (D11). It is important to ensure that PHI and PII data, including, but not limited to, gender and sex-related data, is managed and stored (or shared) in a manner that protects the privacy of participants. De-dentification can be especially challenging for small populations, but this is a topic beyond the scope of this paper. The ethical practitioner is aware of these regulations and assures that their practices are in keeping with policies, regulations, and laws relating to privacy and protecting confidentiality.(Kennedy, 2008; Mayer, 2002)

Responsibilities of the ethical practitioner relating to the privacy of data appear in multiple Ethical Guideline Principles. Some are more implicitly related (e.g., A11: "(the ethical statistical practitioner) Follows applicable policies, regulations, and laws relating to their professional work, unless there is a compelling ethical justification to do otherwise."), while others are explicit (e.g., D; D1; D9). The US law is explicit about assuring the de-identification of otherwise protected data, "De-identified health information neither identifies nor provides a reasonable basis to identify an individual."(DHHS, 2003) There are a myriad of reasons why this information is considered private and protected, see e.g., Wiesemann (2011). Public disclosure of one's sex and/or gender can put gender and sex minorities, especially transgender people, in a position where they are vulnerable to overt discrimination and even to physical violence (e.g., Wirtz et al. (2020), Roffee and Waling (2017), and Jaffee et al. (2016) among other well documented research into gender-based violence). Thus is it pertinent to consider the safety of gender and sex minorities both at the time of data collection (e.g., are respondents asked questions about their gender or sex aloud or by someone who will not be providing them care?) and after data has already been collected. Other published works provide greater detail on the extent of gender-related victimization and discrimination of transgender and gender nonconforming individuals, e.g., Hendricks and Testa (2012).

Additionally, statistical practitioners must rigorously work to ensure that personal information remains confidential. Participants in research or surveys where sensitive data are to be collected should be made aware of extra privacy and confidentiality measures that are to be taken so that they do not feel vulnerable, or pressured to share personal information that they would not ordinarily share/would not want to be exposed. In studies where identifying gender and sex minorities is necessary, participants should be forewarned, protective steps to be taken should be articulated, and the informed consent forms should be as explicit as possible in terms of the intended uses - and potential re-uses - of the data. The consideration of privacy and confidentiality of data is also reflected in the many principles from the ASA guidelines, (e.g., the ethical statistical practitioner "Respects expectations of data contributors when using or sharing data. Exercises due caution to protect proprietary and confidential data, including all data that might inappropriately harm data subjects." (B4); "Understands and conforms to confidentiality requirements for data collection, release, and dissemination and any restrictions on its use established by the data provider (to the extent legally required)."(C7); and "Protects people's privacy and the confidentiality of data concerning them, whether obtained from the individuals directly, other persons, or existing records. Knows and adheres to applicable rules, consents, and guidelines to protect private information."(D4), among others).

The third recommendation from the 2022 article is to center the inclusivity and respect of all research participants. The first two recommendations, ensuring measurement accuracy and specificity, and protecting privacy, are most strongly reflected in both general training and preparation for statistical and data science practice. This is clear from the number and variation in the relevant elements of the Ethical Guidelines for Statistical Practice, which define ". . . 'statistical practice' (to) include activities such as: designing the collection of, summarizing, processing, analyzing, interpreting, or presenting, data; as well as model or algorithm development and deployment.", and which uses ". . . the term 'statistical practitioner' (to) include all those who engage in statistical practice, regardless of job title, profession, level, or field of degree." The third recommendation, centering inclusivity and respect, is qualitatively different because it reflects a more strongly social aspect in terms of encouraging a collaborative relationship between researchers and study subjects than what statistics practitioners may have been trained to expect. Centering inclusivity and respect for sex and gender variables and data implicitly requires purposeful engagement with gender and sex minorities who are directly impacted by (or excluded from) the research. This engagement by researchers with the individuals who are the focus or target of the research maps onto the emphasis on inclusion of patient-reported outcomes (PROs) by the US Food and Drug Administration (see Tractenberg et al. (2017) for a discussion of, and model for, patient-centered PRO vs clinician-centered PRO development).(DHHS-FDA, 2006) Given its divergence from how most statistics practitioners might be trained to engage with collaborators, the remainder of the paper is focused on this third recommendation. In particular, the next section evaluates the extent to which various survey questioning methods attempting to identify gender and sex minorities a) are useful for reproducible and interpretable research; and b) are aligned with ASA Ethical Guidelines for Statistical Practice. In the subsequent section, we explore where centering inclusivity and respect for gender and sex minorities must occur in the tasks that comprise the statistics and data science pipeline (Tractenberg, 2020).

### 3. Moving from connotative categorizations towards descriptive ones

Statisticians are among the best prepared research team members to understand the subjectivity of measurement systems and the impact these have on our statistical conclusions and understanding of variation. However, statistical practitioners may or may not be trained to recognize how this relates to considerations of ethical principles in human research. To attempt to shed some light on this deep relationship between measurement systems and ethical considerations, we consider two different questioning methods.

### 3.1 Two-step questioning method

A hallmark of 21st century research is the inclusion of both gender and sex information in scientific research. The current dominant approach for incorporating gender and sex information is to rely on a two-step questioning procedure where the researcher distinguishes sex and gender. Given the historical treatment of gender and sex in western research, this represents a great improvement in the quality of information (Nielsen et al., 2021). For longitudinal studies in particular, this two-step procedure is consistent with the recognition that gender and gender identity are constructs that can change, and have changed, over time (Hanes and Clouston, 2021). Sex at birth is typically determined by observation of external features apparent at birth. While fast and simple, this method naturally fails to capture differences in sexual development that will occur later in life. (In fact, it is not difficult to find stories of people who did not find out that they are intersex until later in their adult life, e.g. BBC (2020).) If it is important to identify individuals who may have differences in sexual development or who have undergone sex changing surgeries, then the questions for both sex and gender in a longitudinal study should also reflect the potential for distinct sexual characteristics to can change over time.

Whether applied in a longitudinal or cross-sectional context, this two-step questioning procedure captures more human variation than traditional single questions where sex is conflated with gender (Tate et al., 2013). However, depending on how levels within each variable (sex as a biological variable or gender as a socio-cultural variable) are defined, this method may not necessarily respect and include gender and sex minorities in particular.

Consider, for example, the two step questioning procedure recommended by the Committee on Measuring Sex, Gender Identity, and Sexual Orientation to the National Institutes of Health for national and international surveys (NatlAcad, 2022):

Example 3.1.
(Q1) What sex were you assigned at birth, on your original birth certificate? (Female; Male; Don't know; Prefer not to answer)
(Q2) What is your current gender? (Female; Male; Transgender; Two-Spirit; I use a different term:[free text]; Don't know; Prefer not to answer)

Both questions are proposed so respondents can select only one response per question. First, note that this example would require a transgender woman to declare herself as "male" at birth in Q1 - which is basically a report from the delivery room based on the traditional fast and simple physical characteristics, rather than a reflection of her current physical anatomy. Second, this pair of questions does not allow her to unambiguously identify as a woman in Q2, because "transgender" is likely the response intended by the coder or data analyst. That is, if the answer to Q1 is "male" and to Q2 is "woman", the analyst should conclude (based on the comparison, not on the response to Q2 alone) that the individual is a transgender woman - if indeed that information is what is sought. Implicit in these response options is the notion that a transgender person cannot also identify as a woman or a man, and that is unlikely to be the intention of - or even relevant to - the purposes of the data collection.

While better than traditional artificial binary single question method, Example 3.1 violates Principle D6 by failing to consider the impact of these response options on society, groups, and individuals at large within transgender and gender-non-conforming communities. A cisgender respondent may easily (erroneously) infer from the structure of the second question that transgender is something entirely different from identifying as a man or woman. Depending on the response options, this method could end up adversely impacting LGBTQ+ and gender and sex minority groups. Collaboration with community members is important to assure that decisions and inferences based on data collected with a two-step approach will be accurate and representative, supporting B7 while not violating A3. Furthermore, this method is not a universal solution since it is not always the case that human data must include primary sex characteristics. Data collectors should reflect on questions about sex at birth carefully and determine whether or not inducing a potentially severe psychological distress for a minority of respondents (called gender dysphoria - "a concept designated in the DSM-5-TR as clinically significant distress or impairment related to gender incongruence, which may include desire to change primary and/or secondary sex characteristics"(APA, 2022, online)) is necessary for the purpose of the study. Perhaps there is a better way to collect the relevant information that does not put this undue burden on a socially vulnerable sub-population. The questions Example 3.1 may not only cause stress or discomfort for transgender individuals who are being asked to "out" themselves by revealing their sex at birth, but the response options carry heavy informational implications when gender options require participants select one level and distinguish among men, women, and transgender people.

A simple, but powerful, amendment to the example above are the following two questions:

Example 3.2.
(Q1) What sex were you assigned at birth, on your original birth certificate? (Female; Male; Don't know; Prefer not to answer)
(Q2) What is your current gender? (Woman; Man; Non-binary; I use a different term:[free text]; Don't know; Prefer not to answer)

In this amended version of the two-step questioning process, it is now possible for the data collector to identify transgender men (Q1: Female, Q2: Man), transgender women (Q1: Male, Q2: Woman), gender non-conforming (including non-binary) individuals (Q1: Either Male or Female; Q2: Gender non-conforming) as well as cisgender men (Q1: Male; Q2: Men) and cisgender women (Q1: Female, Q2: Woman). Further, it is more likely that "prefer not to answer" truly reflects that preference rather than possibly also including implicit "what I would have chosen was not included in the listed options"(Harrison et al., 2012). Furthermore, recent results from the Williams Institute found that a greater percentage of non-binary LGBTQ adults are cisgender rather than transgender, which can not be reflected in Example 3.1 (Wilson and Meyer, 2021). The amended version avoids unnecessary complication, reflects modern conceptualizations of/scientific interest in gender and sex data, and resolves the issue of potentially conflicting interests whereby a researcher may wish to identify transgender respondents but transgender men and women may wish to assert their gender identity as men and women. It is also obvious from this discussion that if the survey includes any kind of skip pattern, i.e., if you answer "female" to Q1, the second item you see might be different than it is if the answer to Q1 is "male" or (anything else), those patterns must be very carefully mapped out to keep the simplicity of responding as equal as possible for all participants. Whereas some skip pattern might be considered for the first pair of questions (Example 3.1), no skip pattern is intended with Example 3.2 because comparisons of both responses is relevant to generate the categorized characteristic of interest.

Bauer et al. (2017) propose the following two (or three) step questioning method for use in population studies:

Example 3.3.
(Q1) What sex were you assigned at birth, meaning on your original birth certificate? (Male; Female)
(Q2) Which best describes your current gender identity? (Male; Female; Indigenous or other cultural gender minority identity (eg. two-spirit); Something else (e.g. gender fluid, non-binary))
(Q3) What gender do you current live as in your day-to-day life? (Male; Female; Sometimes male, sometimes female; Something other than male or female)[1]

As with Example 3.2, Example 3.3 is completely aligned with Ethical Principle D. Although skip patterns are induced by providing an additional third question to only a few respondents, in this case, the skip pattern is nominal only and can be addressed in a straightforward manner (see footnote).

The two-step questioning method is a great advancement towards meaningfully representing the variation in human characteristics of sex and gender quantitatively (B7). This method is also recommended as a way to avoid exploiting the vulnerable population of gender and sex minorities and to avoid perpetuating unfair outcomes about this group.(A3) Further, two-step questioning is more in-line with the Ethical Guidelines for Professional Integrity and Accountability (Principle A) than a single question procedure that conflates sex and gender. This methodology produces data that are valid, relevant, and appropriate, without favoritism or prejudice, and in a manner intended to produce valid, interpretable, and reproducible results.(A2) By treating gender and sex as separate concepts, this method helps to promote the dignity and fair treatment of all people, including gender and sex minorities. Two-step questioning method also adheres to Principle B pertaining to the Integrity of Data and Methods. When responses for each of the two questions are "select only one", this method clearly communicates details about the data generation and collection processes. It is also inherently transparent about any assumptions made regarding gender and sex so that the analytical conclusions are readily assessed in terms of limitations and possible sources of error or bias.(B1, B2) In large part, the two-step questioning method is a statistical response to understanding what various stakeholders hope to learn from gender and sex data and reflects a better understanding of the subject-matter knowledge necessary to conduct meaningful and relevant statistical analyses of human data.(C1)

---
[1]This question need only asked of those who indicated a current gender identity different than their birth-assigned sex. The information for cisgender individuals can be automatically coded into this third question to avoid any skip patterns from emerging.

## 3.2 Direct questioning method

When possible, direct questioning methods can be used to collect information about gender or sex rather than rely on multiple choice questions. Rather than utilize a more-complex multi-step questioning approach, it may be preferable to ask more simple and direct questions such as "Do you consider yourself transgender?" or "Do you consider yourself to be gender-non-conforming?" A key consideration with this type of question is still ensuring participant safety and comfort during data collection, as well as the usual data storage/availability concerns discussed earlier. More examples of direct questions that capture different dimensions of gender for health research in particular can be found in Tadiri et al. (2021), for example. From a statistical perspective, direct, specifically worded questions reflect careful variable definition from the outset of a study. Given the challenges for the statistical practitioner to assure transparency, and the fitness of data for its intended use, direct questions with yes/no response options may be the best way to get the most informative data in general. If data must be combined or grouped at a later stage in analysis, this can/should be transparently and respectfully implemented. For example, there is no reason for a data curator to avoid longer, descriptive labels for combined groups for, say, "people born with a penis" as opposed to labeling transgender women and cisgender men both in a category called "male".

Asking direct questions with yes/no response options is ethical statistical practice since it enables researchers to capture and describe the variation in human characteristics (B7). This method also avoids some potential challenges arising from the two-step questioning procedure and ensures data is collected in a manner consistent with subjects' consent and it does take into account the broader impact of questionnaires on a societal understanding of sex and gender (D5, D6). Direct questioning methods can be employed to promote the just and impartial treatment of the rights of gender and sex minorities (D11) and, when appropriately used, direct questions with unambiguous yes/no response options meet the highest standards for ethical statistical practice (A2).

In contrast to these other methods of asking for information about sex and gender, some non-binary scientists (DeHority et al., 2021) suggest asking for gender identity using an open-ended question format. Questions like these allow respondents to define their own labels related to sex or gender in answer to a question like, "how do you personally identify your gender?". This method is simple to implement and there has been some positive indication regarding the accuracy of this method, e.g. Fraser et al. (2020). However, it is important to note that the responses from open-ended questions are not necessarily unbiased because respondents and researchers alike may operate under slightly (or dramatically) different definitions of terms that respondents will use in their answers, like "male", "female", or "non-binary". Further, "mostly male" or "male" might have very different meanings for the individuals who responded in these ways, and such responses cannot be categorized confidently. These must be treated subjectively, which can limit the scientific utility of the data point, even if the inclusion of such an item can act to build a sense of community with the respondent. This method does center the inclusivity and respect of gender and sex minorities, but it cannot be recommended as the sole assessment or item for gender data collection and analysis from a statistical perspective. The variability in intention as well as in responses can lead to just as many misclassification errors as an incomplete options list can. Relying entirely on a single open-ended question like this clearly violates ASA Principle A2 because it does not represent "methodology and data that are valid, relevant, and appropriate[...] in a manner intended to produce valid, interpretable, and reproducible results". Wasting resources to complete a study that cannot generate reproducible results is inconsistent with the majority of the Ethical Guidelines for Statistical Practice. Furthermore, as noted by Bauer et al. (2017), "such questions appear to allow participants to avoid simple categorizations, but then participants' identities are categorized by researchers after the fact; the final categorization may be inconsistent with a participant's self-categorization when given those categorical options."

### 3.3 Reflections on these methods

When designing a survey it is crucial for researchers to reflect on what information is really necessary for their study. In fact, the National Academies of Sciences, Engineering, and Medicine Committee on Measuring Sex, Gender Identity, and Sexual Orientation recommends that (the National Institutes of Health in particular) "collect data on gender and report it by default. Collection of data on sex as a biological variable should be limited to circumstances where information about sex traits is relevant... In human populations, collection of data on sex as a biological variable should be accompanied by collection of data on gender."(NatlAcad, 2022, pg. 8) Importantly, a) NIH expectations are informed by its focus on health, where gender might be an important variable ("by default"); b) NIH policy also stipulates that, with appropriate justification, gender and sex data can be omitted from NIH-funded reporting; and c) this NIH language reflects a consideration of "sex" as biological variable whereas gender is treated as social data. Ensuring the clarity of the investigators' purposes in collecting and analyzing gender and sex data enables the ethical practitioner to fulfill many Ethical Guideline elements (throughout Principle B, C, and D). These discussions can also improve transparency and communication well in advance of running any statistical tests or automated selection procedures. This is both a statistical best practice and highly relevant for the safety and comfort of survey respondents. Even in health care settings gender and sex minorities face discriminating and a lack of cultural competence from providers, hindering their access to care. In fact, not all doctors and few medical assistants are educated about transgender identities, experiences, and/or health care (Safer et al., 2016). Careful consideration of whether or not gender or sex information is relevant to the purpose of the data collection is directly related to the consideration of the safety and comfort of gender and sex minority respondents. A transgender women, for example, may be asked these direct questions by various individuals in the routine process of blood testing or registration for the use of a clinic. She may not feel comfortable disclosing that she is transgender out of fear of overt or subtle discrimination. See, for example, results from the focus group responses in Thompson (2016) for real examples of this.

In some cases however, it is necessary to collect information about respondents' sex and gender and these questions must still be carefully considered as they are sensitive for gender and sex minorities. Questions that, when answered truthfully, may result in overt or covert mistreatment of the respondent are not a new issue. For over 50 years, survey designers have had the option of creating a set of questions where the aggregated data reflects an interpretable signal, e.g., an estimate of the proportion of a sample that endorsed a behav-ior or answered "yes" truthfully to a survey question reflecting sensitive information. The survey is designed with direct questions, but rather than being instructed to answer yes/no or choose the response (as shown with the two-step questions), respondents are asked to answer yes, or answer truthfully, only if some circumstance (unknown to the investigator, but pre-established) has been met. Randomized response surveys are designed to introduce a pre-specified amount of noise into survey results such that no individual respondent answer can ever be known (John et al., 2018). This technique has been used in studies about HIV (Arnab and Singh, 2010) and abortion (Perri et al., 2016) for example, to augment protections to the privacy of respondents. Depending on the context of data collection and its availability, randomized response surveys techniques may be an appropriate tool for the study of gender and sex minorities as it protects respondents' information both during and after the study. When used effectively/appropriately (see e.g., John et al. (2018)), this approach can mitigate non-response bias and classification bias that might arise due to safety or identifiablility concerns. While randomized response surveys are a valid statistical technique for collecting sensitive information, the security they provide can be difficult to communicate to participants, resulting in unpredictable noise propagation and unreliable estimates (John et al., 2018).

Irrespective of which question format is used, careful attention must be paid to the questions as well as to any response options provided. For example, asking a question that involves socio-cultural terms such as "transgender" or "gender-non-conforming" without providing a working definition for these terms is subject to misclassification bias. Defining gender and sex-related terms is not a trivial task in science communication and collaboration. As noted earlier, this can benefit from input from stakeholders, research subjects, data analysts, disciplinary experts, and more. The most important structure for the effort of defining terms is to ensure that the purposes for which the data are to be collected are explicit for all stakeholders (C1). This requires a big investment of time, and trust among stakeholders. It may be tempting to skip steps, use existing surveys without consideration of the questions or how they might be perceived or answered. In addition to violating E4 (do not compromise validity for expediency!), the effort, resources, and time expended may ultimately result in data that are simply not fit for purpose. If survey designers fail to provide careful definitions to respondents, data collected from surveys built using any of these methods can contain many different measurement biases as it becomes impossible to tell if the researchers' definition of, say, "masculine" matches that of any given respondent. That misalignment would naturally render the data unfit for purpose (B1). Furthermore, if stakeholders are not identified and consulted before collecting the data on gender or sex, this indicates a failure of the statistical practitioner to obtain sufficient subject-matter knowledge to conduct a meaningful statistical analysis in the first place (C1). Though these questioning methods can be used appropriately, practitioners are obligated to identify potential limitations of statistical methods and we would be remiss to fail to acknowledge the potentially high costs involved with the different methods of survey design (C4).

4. Ethical considerations throughout the statistics and data science pipeline

The Statistics and Data Science Pipeline, defined by Tractenberg (2020), identifies seven general tasks that those who work with data (applying statistical practice or doing data science) engage in and/or recognize:

1. Planning/Designing;

2. Data collection/munging/wrangling;

3. Analysis (perform or program to perform);

4. Interpretation;

5. Documenting your work;

6. Reporting your results/communication;

7. Engaging in team work.

The first two stages concern the creation or curation of usable data, tasks three and four are concerned with understanding the signal(s) in the data, and the last three tasks deal with communication. Consideration of the storage and/or sharing of data arise in every single task. Assuring measurement is meaningful (recommendation 1) is principally the focus of tasks 1-2, but task 5-6 (documenting/reporting) are also key for communicating to others what the data purport to measure. Privacy is a core feature in tasks 1-2, but protections for privacy can also arise in the documentation of the work and other communication (tasks 5-6). Opportunities to center the inclusivity and respect for research subjects can arise, or be created, at every one of these tasks.

## 4.1 Tasks 1-2: Planning/Designing and Data collection/munging/wrangling

Statisticians are well-aware that the scope of our analytic conclusions is inextricably linked to the way in which data is generated, curated, or collected. As discussed in the previous section, data collection strategies may or may not enable researchers to respect and include gender and sex minorities. In any case, the variables of interest for a particular study must be carefully considered and clearly defined for any statistical analysis to be meaningful. This means researchers (both data collectors and analyzers) must be able to answer fundamental questions such as "What information is being measured or counted?"; "Why is this information important for my research question(s)?"; and "What could bias my measurement of this information?" The recommendation for scientists to center the inclusivity and respect of gender and sex minorities in their research automatically invokes this deeper critical thinking about gender and sex variables. Acknowledging and respecting the experience and identities of transgender and intersex individuals immediately illuminates the limitations of typical binary categorizations of gender and sex. Once these limitations are acknowledged, the researcher must justify how to proceed, prompting these fundamental questions about gender and sex in particular. For biomedical research on sex differences Rich-Edwards et al. (2018) rightly note that "[w]ithout careful methodology, the pursuit of sex difference research, despite a mandate from funding agencies, will result in a literature of contradiction." The same is also true generally for human sciences that seek to understand or identify sex and/or gender differences, and to incorporate these differences into the studies of other human characteristics and outcomes. It is important to keep in mind that plans and collection efforts may be done by teams or individuals other than those who will be executing other tasks. The Ethical Guidelines recommend a variety of considerations for fellow statistics practitioners (Principle F), but also entail many different elements encouraging transparency (B2, B3, E3), sharing data as is possible (B4, D5, D7, F4), and complete and accurate documentation of the work (D10, F4).

Moreover, storing, managing, and sharing (if relevant) of data should be contemplated as part of the planning and collecting tasks. Human studies require IRB approval which addresses, in part, protections to research subjects' privacy once data is in storage. The Department of Homeland Security, for example, provides guidelines for safeguarding personally identifiable information.(DHS, 2017) However, depending on local and federal political climates, gender and sex minority participants may or may not be protected against overt discrimination. Participant privacy is crucial for data that identifies gender or sex minorities both because of a lack of legal protections and because this group is vulnerable to social discrimination, harassment, and often violence. (See earlier sources for examples.) If the data is meant to be used again in the future and possibly shared with others, determining and addressing any identifiability concerns is a necessary step for ensuring the safety and privacy of gender and sex minority study participants. If the participants are identifiable, then the data cannot be shared without their explicit permission. Furthermore, the ASA Ethical Guidelines articulate an obligation for statistical practitioners to "not knowingly conduct statistical practices that exploit vulnerable populations or create or perpetuate unfair outcomes" (A3); stating that the ethical practitioner "Respects expectations of data contributors when using or sharing data. Exercises due caution to protect proprietary and confidential data, including all data that might inappropriately harm data subjects" (B4); "Understands and conforms to confidentiality requirements for data collection, release, and dissemination and any restrictions on its use established by the data provider (to the extent legally required). Protects the use and disclosure of data accordingly. Safeguards privileged information of the employer, client, or funder." (C7); "Protects people's privacy and the confidentiality of data concerning them, whether obtained from the individuals directly, other persons, or existing records. Knows and adheres to applicable rules, consents, and guidelines to protect private information." (D4); "Knows the legal limitations on privacy and confidentiality assurances and does not over-promise or assume legal privacy and confidentiality protections where they may not apply." (D9).

## 4.2 Tasks 3-4: Analysis (perform or program to perform) and Interpretation

Once cleaned data is in hand, the next tasks are to analyze the data. Analyses often include exploratory visualizations and numeric summaries or inferential statements or both. The quality and nature of the statistical conclusions possible from the analysis stage are inextricably linked to the manner in which the data was collected. Thus it is always recommended to plan analyses before gathering data, so in some sense, the data analysis stage should partly occur before the data collection stage. In the planning stage of analyses, many potential issues with both data collection and variable definition can be identified and mitigated to conserve resources and ensure relevant data is collected. Small sample size is a common issue for analyses that need to detect small human sub-populations, including gender and sex minorities. A pre-planned analysis can identify the sample size issue early on and sampling strategies can be integrated into the data collection procedure in a way that avoids biasing inferential conclusions. The pre-planning can also enable protections against unethical statistical practices (Wang et al., 2018) and promotes compliance with multiple Ethical Guideline Principles (A, B, C, E and F).

The common, but often avoidable, situation where small sample representation of minority sub-populations occurs tends to result in reducing the dimension of categorical data. In extreme examples, quantitative variables, like age, may be reduced to ordinal categorical levels corresponding to key developmental stages. Sometimes this dimension-reduction solution is unavoidable. In the case of sex and gender related variables, the reduction can still be implemented in a way that centers the inclusivity and respect of gender and sex minorities. For example, if data was collected on participants' gender with three levels: man, woman, or gender-non-conforming, it is very likely that the sample size of gender-non-conforming individuals may be too small to work with in any inferential analyses. In this situation, the researcher must make a decision regarding the treatment of the third level. Solutions may be to group it together with one of the other levels, split it in half between the two other categories, and/or try all of these methods and compare the results. Whenever re-leveling variables related to gender, researchers should take care to avoid mis-gendering participants and should include a transparent discussion on the methods of and reasoning behind the re-leveling process.

Finally, consider a comment on longitudinal studies where gender or sex are variables of interest. Researchers should note that one's gender is not necessarily fixed in time. (Depending on how sex is defined, this could also be the case with sex as a variable of interest.) The ASA Ethical Guidelines note that the ethical statistical practitioner "Explores and describes the effect of variation in human characteristics and groups on statistical practice when feasible and relevant", (B7) but this must be done in keeping with methodological validity (A2), and without harming others ("Does not knowingly conduct statistical practices that exploit vulnerable populations or create or perpetuate unfair outcomes" (A3); Moreover, "The ethical statistical practitioner does not misuse or condone the misuse of data. They protect and respect the rights and interests of human and animal subjects. These responsibilities extend to those who will be directly affected by statistical practices." (Principle D)

### 4.3 Tasks 5-6: Documenting your work and Reporting your results/communication

Once analyses are completed and the target inferences are drawn, what was done should be documented and results should be shared. The documentation and communication should be completed in a manner that is consistent with the Ethical Guidelines (e.g., D10, D11; E3, E4; F4). At this point in a project, whether you are the team/individual that executed the other tasks as well, or you are just tasked with writing the documentation, it is important to represent the methods that were used as coherently and completely as possible so as to facilitate the rigor and reproducibility of future researchers who use the data, methods, or results you are documenting and reporting. Briefly, it is also important the center the inclusivity of gender and sex minorities in the final data pipeline stage of engaging teamwork. Human characteristics like gender and sex are not only subjects of analyses, but are real and meaningful parts of any individual's sense of self, whether that individual is a researcher or a data subject. For effective team work, one should not assume know another team member's identity or personal experience with sex and gender.

## 5. Discussion and conclusion

Modern biomedical and social science research standards are finally recognizing the severe limitations of an artifactual, binary categorization of sex and gender. The reliance on traditional, overly simplistic, binary classification of sex and gender should not be allowed to limit scientific and social advances. Safe and informative scientific studies can be designed when individual identities and experiences are respected. Studies designed with more appropriate representation of sex and gender will be more rigorous, more reproducible, and more inclusive than those without these characteristics. The considerations discussed in this paper are not unique to variables representing sex and gender, nor to decisions that are informed by analyses involving these variables. It is straightforward to apply the survey design choice discussions to other variables that are similarly subject to a history of artifactual and uninformative binary categorizations (e.g., race, ethnicity). Consideration of the relevance of aspects of the ethical practice standards for statistics and data science to sex and gender variables is timely, with results that can be applied more widely to other variable definition and selection habits that tend to prioritize expediency ("it's what we've always done!") over validity, rigor, and reproducibility. Clear and effective science communication and widespread statistical literacy emerge as related endeavors.

Core elements of the ASA Ethical Guidelines that always pertain, but are particularly important when considering sex and gender variables are that the ethical statistical practitioner:

- A2 Uses methodology and data that are valid, relevant, and appropriate, without favoritism or prejudice, and in a manner intended to produce valid, interpretable, and reproducible results.

- A3 Does not knowingly conduct statistical practices that exploit vulnerable populations or create or perpetuate unfair outcomes.

- A11 Follows applicable policies, regulations, and laws relating to their professional work, unless there is a compelling ethical justification to do otherwise.

- B......seeks to understand and mitigate known or suspected limitations, defects, or biases in the data or methods and communicates potential impacts on the interpretation, conclusions, recommendations, decisions, or other results of statistical practices.

- B2 Is transparent about assumptions made in the execution and interpretation of statistical practices including methods used, limitations, possible sources of error, and algorithmic biases. Conveys results or applications of statistical practices in ways that are honest and meaningful.

- B7 Explores and describes the effect of variation in human characteristics and groups on statistical practice when feasible and relevant.

- C1 Seeks to establish what stakeholders hope to obtain from any specific project. Strives to obtain sufficient subject-matter knowledge to conduct meaningful and relevant statistical practice.

- C2 Regardless of personal or institutional interests or external pressures, does not use statistical practices to mislead any stakeholder.

- D........does not misuse or condone the misuse of data. They protect and respect the rights and interests of human and animal subjects. These responsibilities extend to those who will be directly affected by statistical practices.

- D9 Knows the legal limitations on privacy and confidentiality assurances and does not over-promise or assume legal privacy and confidentiality protections where they may not apply.

- D10 Understands the provenance of the data, including origins, revisions, and any restrictions on usage, and fitness for use prior to conducting statistical practices.

- D11 Does not conduct statistical practice that could reasonably be interpreted by subjects as sanctioning a violation of their rights. Seeks to use statistical practices to promote the just and impartial treatment of all individuals.

- E4 Avoids compromising validity for expediency. Regardless of pressure on or within the team, does not use inappropriate statistical practices.

- F4 Promotes reproducibility and replication, whether results are "significant" or not, by sharing data, methods, and documentation to the extent possible.

According to modern ethical standards, especially those identified in the Belmont Report, it is crucial that our research about human subjects respects the subjects' sense of self and self-identity. This is further emphasized for statisticians and data scientists throughout the ASA Ethical Guidelines for Statistical Practice. Whatever their discipline, job title, or training, statistical practitioners should carefully consider the way sex and gender information is coded in the data sets they design, curate, collect, or analyze/study. If biological factors related to reproductive roles are important, think more critically about why they might be important, what features in particular play a role? What about people who are infertile? What about those who know this vs those who don't? Ethical principle B7 calls for the use of statistical methods that accurately represent the variability inherent to a population. In addition to asking the right questions, researchers must communicate the limitations of their measurement systems and consider the well-being of their research subjects as individuals and as members of different communities. The effects of gender and sex variation among humans is challenging to study but it is impossible to develop a scientific understanding of these effects if we aren't asking the right questions or using clear terminology.

## Acknowledgements

ST acknowledges support from Swarthmore College Faculty Research Support Grants. RET has none.

# Appendix

## Ethical Guidelines for Statistical Practice

Prepared by the Committee on Professional Ethics of the American Statistical Association on 31 January 2022.

PURPOSE OF THE GUIDELINES:

The American Statistical Association's Ethical Guidelines for Statistical Practice are intended to help statistical practitioners make decisions ethically. In these Guidelines, "statistical practice" includes activities such as: designing the collection of, summarizing, processing, analyzing, interpreting, or presenting, data; as well as model or algorithm development and deployment. Throughout these Guidelines, the term "statistical practitioner" includes all those who engage in statistical practice, regardless of job title, profession, level, or field of degree. The Guidelines are intended for individuals, but these principles are also relevant to organizations that engage in statistical practice.

The Ethical Guidelines aim to promote accountability by informing those who rely on any aspects of statistical practice of the standards that they should expect. Society benefits from informed judgments supported by ethical statistical practice. All statistical practitioners are expected to follow these Guidelines and to encourage others to do the same.

In some situations, Guideline principles may require balancing of competing interests. If an unexpected ethical challenge arises, the ethical practitioner seeks guidance, not exceptions, in the Guidelines. To justify unethical behaviors, or to exploit gaps in the Guidelines, is unprofessional, and inconsistent with these Guidelines.

A   PRINCIPLE A: Professional Integrity and Accountability

Professional integrity and accountability require taking responsibility for one's work. Ethical statistical practice supports valid and prudent decision making with appropriate methodology. The ethical statistical practitioner represents their capabilities and activities honestly, and treats others with respect.

The ethical statistical practitioner:

1. Takes responsibility for evaluating potential tasks, assessing whether they have (or can attain) sufficient competence to execute each task, and that the work and timeline are feasible. Does not solicit or deliver work for which they are not qualified, or that they would not be willing to have peer reviewed.

2. Uses methodology and data that are valid, relevant, and appropriate, without favoritism or prejudice, and in a manner intended to produce valid, interpretable, and reproducible results.

3. Does not knowingly conduct statistical practices that exploit vulnerable populations or create or perpetuate unfair outcomes.

4. Opposes efforts to predetermine or influence the results of statistical practices, and resists pressure to selectively interpret data.

5. Accepts full responsibility for their own work; does not take credit for the work of others; and gives credit to those who contribute. Respects and acknowledges the intellectual property of others.

6. Strives to follow, and encourages all collaborators to follow, an established protocol for authorship. Advocates for recognition commensurate with each person's contribution to the work. Recognizes that inclusion as an author does imply, while acknowledgement may imply, endorsement of the work.

7. Discloses conflicts of interest, financial and otherwise, and manages or resolves them according to established policies, regulations, and laws.

8. Promotes the dignity and fair treatment of all people. Neither engages in nor condones discrimination based on personal characteristics. Respects personal boundaries in interactions and avoids harassment including sexual harassment, bullying, and other abuses of power or authority.

9. Takes appropriate action when aware of deviations from these Guidelines by others.

10. Acquires and maintains competence through upgrading of skills as needed to maintain a high standard of practice.

11. Follows applicable policies, regulations, and laws relating to their professional work, unless there is a compelling ethical justification to do otherwise.

12. Upholds, respects, and promotes these Guidelines. Those who teach, train, or mentor in statistical practice have a special obligation to promote behavior that is consistent with these Guidelines.

B    PRINCIPLE B: Integrity of Data and Methods

The ethical statistical practitioner seeks to understand and mitigate known or suspected limitations, defects, or biases in the data or methods and communicates potential impacts on the interpretation, conclusions, recommendations, decisions, or other results of statistical practices.

The ethical statistical practitioner:

1. Communicates data sources and fitness for use, including data generation and collection processes and known biases. Discloses and manages any conflicts of interest relating to the data sources. Communicates data processing and transformation procedures, including missing data handling.

2. Is transparent about assumptions made in the execution and interpretation of statistical practices including methods used, limitations, possible sources of error, and algorithmic biases. Conveys results or applications of statistical practices in ways that are honest and meaningful.

3. Communicates the stated purpose and the intended use of statistical practices. Is transparent regarding a priori versus post hoc objectives and planned versus unplanned statistical practices. Discloses when multiple comparisons are conducted, and any relevant adjustments.

4. Meets obligations to share the data used in the statistical practices, for example, for peer review and replication, as allowable.

5. Respects expectations of data contributors when using or sharing data. Exercises due caution to protect proprietary and confidential data, including all data that might inappropriately harm data subjects.

6. Strives to promptly correct substantive errors discovered after publication or implementation. As appropriate, disseminates the correction publicly and/or to others relying on the results.

7. For models and algorithms designed to inform or implement decisions repeatedly, develops and/or implements plans to validate assumptions and assess performance over time, as needed. Considers criteria and mitigation plans for model or algorithm failure and retirement.

8. Explores and describes the effect of variation in human characteristics and groups on statistical practice when feasible and relevant.

C    PRINCIPLE C: Responsibilities to Stakeholders

Those who fund, contribute to, use, or are affected by statistical practices are considered stakeholders. The ethical statistical practitioner respects the interests of stakeholders while practicing in compliance with these Guidelines.

The ethical statistical practitioner:

1. Seeks to establish what stakeholders hope to obtain from any specific project. Strives to obtain sufficient subject-matter knowledge to conduct meaningful and relevant statistical practice.

2. Regardless of personal or institutional interests or external pressures, does not use statistical practices to mislead any stakeholder.

3. Uses practices appropriate to exploratory and confirmatory phases of a project, differentiating findings from each so the stakeholders can understand and apply the results.

4. Informs stakeholders of the potential limitations on use and re-use of statistical practices in different contexts and offers guidance and alternatives, where appropriate, about scope, cost, and precision considerations that affect the utility of the statistical practice.

5. Explains any expected adverse consequences from failing to follow through on an agreed-upon sampling or analytic plan.

6. Strives to make new methodological knowledge widely available to provide benefits to society at large. Presents relevant findings, when possible, to advance public knowledge.

7. Understands and conforms to confidentiality requirements for data collection, release, and dissemination and any restrictions on its use established by the data provider (to the extent legally required). Protects the use and disclosure of data accordingly. Safeguards privileged information of the employer, client, or funder.

8. Prioritizes both scientific integrity and the principles outlined in these Guidelines when interests are in conflict.

D  PRINCIPLE D: Responsibilities to Research Subjects, Data Subjects, or those directly affected by statistical practices

The ethical statistical practitioner does not misuse or condone the misuse of data. They protect and respect the rights and interests of human and animal subjects. These responsibilities extend to those who will be directly affected by statistical practices.

The ethical statistical practitioner:

1. Keeps informed about and adheres to applicable rules, approvals, and guidelines for the protection and welfare of human and animal subjects. Knows when work requires ethical review and oversight.[2]

2. Makes informed recommendations for sample size and statistical practice methodology in order to avoid the use of excessive or inadequate numbers of subjects and excessive risk to subjects

3. For animal studies, seeks to leverage statistical practice to reduce the number of animals used, refine experiments to increase the humane treatment of animals, and replace animal use where possible. Protects people's privacy and the confidentiality of data concerning them, whether obtained from the individuals directly, other persons, or existing records. Knows and adheres to applicable rules, consents, and guidelines to protect private information.

---

[2]Examples of ethical review and oversight include an Institutional Review Board (IRB), an Institutional Animal Care and Use Committee (IACUC), or a compliance assessment.

4. Uses data only as permitted by data subjects' consent when applicable or considering their interests and welfare when consent is not required. This includes primary and secondary uses, use of repurposed data, sharing data, and linking data with additional data sets.

5. Considers the impact of statistical practice on society, groups, and individuals. Recognizes that statistical practice could adversely affect groups or the public perception of groups, including marginalized groups.

6. Considers approaches to minimize negative impacts in applications or in framing results in reporting.

7. Refrains from collecting or using more data than is necessary. Uses confidential information only when permitted and only to the extent necessary. Seeks to minimize the risk of re-identification when sharing de-identified data or results where there is an expectation of confidentiality. Explains any impact of de-identification on accuracy of results.

8. To maximize contributions of data subjects, considers how best to use available data sources for exploration, training, testing, validation, or replication as needed for the application. The ethical statistical practitioner appropriately discloses how the data are used for these purposes and any limitations.

9. Knows the legal limitations on privacy and confidentiality assurances and does not over-promise or assume legal privacy and confidentiality protections where they may not apply.

10. Understands the provenance of the data, including origins, revisions, and any restrictions on usage, and fitness for use prior to conducting statistical practices.

11. Does not conduct statistical practice that could reasonably be interpreted by subjects as sanctioning a violation of their rights. Seeks to use statistical practices to promote the just and impartial treatment of all individuals.

E   PRINCIPLE E: Responsibilities to members of multidisciplinary teams

Statistical practice is often conducted in teams made up of professionals with different professional standards. The statistical practitioner must know how to work ethically in this environment.

The ethical statistical practitioner:

1. Recognizes and respects that other professions may have different ethical standards and obligations. Dissonance in ethics may still arise even if all members feel that they are working towards the same goal. It is essential to have a respectful exchange of views.

2. Prioritizes these Guidelines for the conduct of statistical practice in cases where ethical guidelines conflict.

3. Ensures that all communications regarding statistical practices are consistent with these Guidelines. Promotes transparency in all statistical practices.

4. Avoids compromising validity for expediency. Regardless of pressure on or within the team, does not use inappropriate statistical practices.

## F PRINCIPLE F: Responsibilities to Fellow Statistical Practitioners and the Profession

Statistical practices occur in a wide range of contexts. Irrespective of job title and training, those who practice statistics have a responsibility to treat statistical practitioners, and the profession, with respect. Responsibilities to other practitioners and the profession include honest communication and engagement that can strengthen the work of others and the profession.

The ethical statistical practitioner:

1. Recognizes that statistical practitioners may have different expertise and experiences, which may lead to divergent judgments about statistical practices and results. Constructive discourse with mutual respect focuses on scientific principles and methodology and not personal attributes.

2. Helps strengthen, and does not undermine, the work of others through appropriate peer review or consultation. Provides feedback or advice that is impartial, constructive, and objective.

3. Takes full responsibility for their contributions as instructors, mentors, and supervisors of statistical practice by ensuring their best teaching and advising – regardless of an academic or non-academic setting – to ensure that developing practitioners are guided effectively as they learn and grow in their careers.

4. Promotes reproducibility and replication, whether results are "significant" or not, by sharing data, methods, and documentation to the extent possible.

5. Serves as an ambassador for statistical practice by promoting thoughtful choices about data acquisition, analytic procedures, and data structures among non-practitioners and students. Instills appreciation for the concepts and methods of statistical practice.

## G PRINCIPLE G: Responsibilities of Leaders, Supervisors, and Mentors in Statistical Practice

Statistical practitioners leading, supervising, and/or mentoring people in statistical practice have specific obligations to follow and promote these Ethical Guidelines. Their support for – and insistence on – ethical statistical practice are essential for the integrity of the practice and profession of statistics as well as the practitioners themselves.

Those leading, supervising, or mentoring statistical practitioners are expected to:

1. Ensure appropriate statistical practice that is consistent with these Guidelines. Protect the statistical practitioners who comply with these Guidelines, and advocate for a culture that supports ethical statistical practice.

2. Promote a respectful, safe, and productive work environment. Encourage constructive engagement to improve statistical practice. Identify and/or create opportunities for team members/mentees to develop professionally and maintain their proficiency.

3. Advocate for appropriate, timely, inclusion and participation of statistical practitioners as contributors/collaborators. Promote appropriate recognition of the contributions of statistical practitioners, including authorship if applicable.

4. Establish a culture that values validation of assumptions, and assessment of model/algorithm performance over time and across relevant subgroups, as needed. Communicate with relevant stakeholders regarding model or algorithm maintenance, failure, or actual or proposed modifications.

## H  PRINCIPLE H: Responsibilities Regarding Potential Misconduct

The ethical statistical practitioner understands that questions may arise concerning potential misconduct related to statistical, scientific, or professional practice. At times, a practitioner may accuse someone of misconduct, or be accused by others. At other times, a practitioner may be involved in the investigation of others' behavior. Allegations of misconduct may arise within different institutions with different standards and potentially different outcomes. The elements that follow relate specifically to allegations of statistical, scientific, and professional misconduct.

The ethical statistical practitioner:

1. Knows the definitions of, and procedures relating to, misconduct in their institutional setting. Seeks to clarify facts and intent before alleging misconduct by others. Recognizes that differences of opinion and honest error do not constitute unethical behavior.

2. Avoids condoning or appearing to condone statistical, scientific, or professional misconduct. Encourages other practitioners to avoid misconduct or the appearance of misconduct.

3. Does not make allegations that are poorly founded, or intended to intimidate. Recognizes such allegations as potential ethics violations.

4. Lodges complaints of misconduct discreetly and to the relevant institutional body. Does not act on allegations of misconduct without appropriate institutional referral, including those allegations originating from social media accounts or email listservs.

5. Insists upon a transparent and fair process to adjudicate claims of misconduct. Maintains confidentiality when participating in an investigation. Discloses the investigation results honestly to appropriate parties and stakeholders once they are available.

6. Refuses to publicly question or discredit the reputation of a person based on a specific accusation of misconduct while due process continues to unfold.

7. Following an investigation of misconduct, supports the efforts of all parties involved to resume their careers in as normal a manner as possible, consistent with the outcome of the investigation.

8. Avoids, and acts to discourage, retaliation against or damage to the employability of those who responsibly call attention to possible misconduct.

## References

(2022). Nature journals raise the bar on sex and gender reporting in research. *Nature*, 605(396).

Agresti, A. (2003). *Categorical Data Analysis*. John Wiley & Sons, Inc., Hoboken, New Jersey.

APA (2022). What is gender dysphoria? www.psychiatry.org/patients-families/gender-dysphoria/what-is-gender-dysphoria. American Psychiatric Association.

Arnab, R. and Singh, S. (2010). Randomized response techniques: An application to the botswana aids impact survey. *Journal of Statistical Planning and Inference*, 140(4):941–953.

Bauer, G., Braimoh, J., Scheim, A., and Dharma, C. (2017). Transgender-inclusive measures of sex/gender for population surveys: Mixed-methods evaluation and recommendations. *PLOS ONE*, 12(5).

BBC (2020). Woman discovered she is intersex in her 40s. www.bbc.com/news/uk-england-humber-51494730. BBC News.

Beery, A. and Zucker, I. (2011). Sex bias in neuroscience and biomedical research. *Neuroscience & Biobehavioral Reviews*, 35(3):565–572.

DeHority, R., Baez, R., Burnette, R., and Howell, L. (2021). Non-binary scientists want funding agencies to change how they collect gender data. *Scientific American*. Ethics: Opinion.

DHHS (2003). Summary of the HIPAA privacy rule. www.hhs.gov/sites/default/files/privacysummary.pdf. U. S. Department of Health and Human Services.

DHHS-FDA (2006). Guidance for industry: Patient-reported outcome measures: Use in medical product development to support labeling claims: Draft guidance. *Health and Quality of Life Outcomes*, 4(79). U.S. Department of Health and Human Services Food and Drug Administration.

DHS (2017). Handbook for safeguarding sensitive pii, privacy policy directive 047-01-007 (Revision 3). Department of Homeland Security Privacy Office of the United States.

Fraser, G., Bulbulia, J., Greaves, L., Wilson, M., and Sibley, C. (2020). Coding responses to an open-ended gender measure in a new zealand national sample. *The Journal of Sex Research*, 57(8):979–986.

Hanes, D. and Clouston, S. (2021). Ask again: Including gender identity in longitudinal studies of aging. *The Gerontologist*, 61(5):640–649.

Harrison, J., Grant, J., and Herman, J. (2012). A gender not listed here: Genderqueers, gender rebels, and otherwise in the national transgender discrimination survey. *LGBTQ Public Policy Journal*, 2:13–24.

Heidari, S., T.F.Babor, DeCastro, P., Tort, S., and Curno, M. (2016). Sex and gender equity in research: Rationale for the sager guidelines and recommended use. *Research Integrity and Peer Review*, 1(2).

Hendricks, M. and Testa, R. (2012). A conceptual framework for clinical work with transgender and gender nonconforming clients: An adaptation of the minority stress model. *Professional Psychology: Research and Practice*, 43(5):460–467.

HIPPA (2022). What is considered protected health information under HIPAA? www.hipaajournal.com/what-is-considered-protected-health-information-under-hipaa. 2014-2022 HIPAA Journal.

Holdcroft, A. (2007). Integrating the dimensions of sex and gender into basic life sciences research: Methodologic and ethical issues. *Gender Medicine*, 4(Suppl B):S64–S74.

Jaffee, K., Shires, D., and Stroumsa, D. (2016). Discrimination and delayed health care among transgender women and men: Implications for improving medical education and health care delivery. *Medical Care*, 54(11):1010–1016.

John, L., Loewenstein, G., Acquisti, A., and Vosgerau, J. (2018). When and why randomized response techniques (fail to) elicit the truth. *Organizational Behavior and Human Decision Processes*, 148:101–123.

Kachel, S., Steffens, M., and Niedlich, C. (2016). Traditional masculinity and femininity: Validation of a new scale assessing gender roles. *Frontiers in Psychology*, 7.

Kennedy, J. (2008). Confidentiality. In Lavrakas, P. J., editor, *Encyclopedia of Survey Research Methods*, pages 132–132. Sage Publications, Inc.

Larsen, R. J. and Marx, M. L. (1986). *An Introduction to Mathematical Statistics and its Applications*. Prentice-Hall. (2nd Edition).

Mayer, T. (2002). Privacy and confidentiality research and the u.s. census bureau recommendations based on a review of the literature. www.census.gov/content/dam/Census/library/working-papers/2002/adrm/rsm2002-01.pdf.

McCarthy, M., Arnold, A., Ball, G., Blaustein, J., and Vries, G. (2012). Sex differences in the brain: The not so inconvenient truth. *Journal of Neuroscience*, 32(7):2241–2247.

McGregor, A., Hasnain, M., Sandberg, K., Morrison, M., Berlin, M., and Trott, J. (2016). How to study the impact of sex and gender in medical research: A review of resources. *Biology of Sex Differences*, 7(Suppl 1)(46).

Miller, V., Kaplan, J., Schork, N., Ouyang, P., Berga, S., Wenger, N., Shaw, L., Webb, R., Mallampalli, M., Steiner, M., Taylor, D. A., Merz, C., and Reckelhoff, J. (2011). Strategies and methods to study sex differences in cardiovascular structure and function: A guide for basic scientists. *Biology of Sex Differences*, 2:14.

NatlAcad (2022). Measuring sex, gender identity, and sexual orientation. The National Academies Press, Washington, D.C. National Academies of Sciences, Engineering, and Medicine.

Nielsen, M., Stefanick, M., Peragine, D., Neilands, T., Ioannidis, J., Pilote, L., Prochaska, J., Cullen, M., Einstein, G., Klinge, I., LeBlanc, H., Paik, H., and Schiebinger, L. (2021). Gender-related variables for health research. *Biology of Sex Differences*, 12(1):1–16.

Perri, P., Pelle, E., and Stranges, M. (2016). Estimating induced abortion and foreign irregular presence using the randomized response crossed model. *Social Indicators Research*, 129:601–618.

Rich-Edwards, J., Kaiser, U., Chen, G., Manson, J., and Goldstein, J. (2018). Sex and gender differences research design for basic, clinical, and population studies: Essentials for investigators. *Endocrine Reviews*, 39(4):424–439.

Ritchie, I. (2003). Sex tested, gender verified: Controlling female sexuality in the age of containment. *Sports History Review*, 34(1):80–93.

Roffee, J. and Waling, A. (2017). Resolving ethical challenges when researching with minority and vulnerable populations: Lgbtiq victims of violence, harassment and bullying. *Research Ethics*, 13(1):4—-22.

Safer, J., Coleman, E., Feldman, J., Garofalo, R., Hembree, W., Radix, A., and Sevelius, J. (2016). Barriers to healthcare for transgender individuals. *Current Opinion in Endocrinology, Diabetes, and Obesity*, 23(2):168–171.

Tadiri, C., Raparelli, V., Abrahamowicz, M., Kautzy-Willer, A., Kublickiene, K., Herrero, M., Norris, C., and Pilote, L. (2021). Methods for prospectively incorporating gender into health sciences research. *Journal of Clinical Epidemiology*, 129:191–197.

Tate, C., Ledbetter, J., and Youssef, C. (2013). A two-question method for assessing gender categories in the social and medical sciences. *Journal of Sex Research*, 50(8):767–776.

Thompson, H. (2016). Patient perspectives on gender identity data collection in electronic health records: An analysis of disclosure, privacy, and access to care. *Transgender Health*, 1(1):205–215.

Thornton, S., Roy, D., Parry, S., LaLonde, D., Martinez, W., Ellis, R., and Corliss, D. (2022). Towards best practices for collecting gender and sex data. *Significance*, 19(1):40–45.

Tractenberg, R. (2020). Ten simple rules for integrating ethics into statistics and data science instruction. *SocArXiv*, 15.

Tractenberg, R., Garver, A., Ljungberg, I., Schladen, M., and Groah, S. (2017). Maintaining primacy of the patient perspective in the development of patient-centered patient reported outcomes. *PLOS ONE*, 12(3).

Wackwitz, L. (2003). Verifying the myth: Olympic sex testing and the category 'woman'. *Women's Studies International Forum*, 26(6):553–560.

Wang, M., Yan, A., and Katz, R. (2018). Researcher requests for inappropriate analysis and reporting: A us survey of consulting biostatisticians. *Annals of Internal Medicine*, 169(8):554–558.

Wiesemann, C. (2011). Is there a right not to know one's sex? the ethics of 'gender verification' in women's sports competition. *Journal of Medical Ethics*, 37(4):216–220.

Wilson, B. and Meyer, I. (2021). Nonbinary lgbtq adults in the united states. *UCLA School of Law: The Williams Institute*. Accessed 28 September 2022.

Wirtz, A., Poteat, T., Malik, M., and Glass, N. (2020). Gender-based violence against transgender people in the united states: A call for research and programming. *Trauma, Violence, & Abuse*, 21(2):227–241.